\documentclass[twocolumn,showpacs,preprintnumbers,amsmath,amssymb,floatfix]{revtex4}

\usepackage{graphicx}
\usepackage{dcolumn}
\usepackage{bm}

\newcommand{\pom}{\tt I\! P}
\newcommand{\beq}{\begin{equation}}
\newcommand{\eeq}{\end{equation}}

\begin{document}

\title{Estimate of the single diffractive heavy quark production in heavy ion collisions at the CERN-LHC}


\author{M. B. Gay Ducati$^a$, M. M. Machado$^a$}
\author{M. V. T. Machado$^b$}

\affiliation{$^a$ High Energy Physics Phenomenology Group, GFPAE,  IF-UFRGS \\
Caixa Postal 15051, CEP 91501-970, Porto Alegre, RS, Brazil}
\affiliation{$^b$ Universidade Federal do Pampa. Centro de Ci\^encias Exatas e Tecnol\'ogicas, \\
Campus de Bag\'e, Rua Carlos Barbosa. CEP 96400-970. Bag\'e, RS, Brazil}

\begin{abstract}

The single diffractive cross section for heavy quarks production is calculated in next-no-leading order (NLO)  for  nucleus-nucleus collisions. Such  processes are expected to occur at the LHC, where the nuclei involved are lead at $\sqrt{s}=$ 5.5 TeV and calcium at $\sqrt{s}=$ 6.3 TeV. We start using the hard diffractive factorization formalism, taking into account a recent experimental parameterization for the Pomeron Structure Function (DPDF). Absorptive corrections are accounted by  the multiple Pomeron contributions considering a gap survival probability, where its theoretical uncertainty for nuclear collisions is discussed. We estimate the diffractive ratios for single diffraction process in nuclear coherent/incoherent collisions at the LHC.
\end{abstract}


\maketitle

\section{Introduction}

Over the last years, a number of high energy hard diffractive processes have used to shed light on our knowledge about the QCD Pomeron. One of them is the diffractive heavy quark production. Central diffractive heavy quarks are quite important signals of possible new physics, for instance in diffractive Higgs production, $H\rightarrow b\bar{b}$ and $WH\rightarrow W\,b\bar{b}$. The knowledge of the characteristic final states with charm or bottom quarks is fundamental to learn how to isolate signals from background for very specific production mechanisms, and also, selects specific classes of relevant higher-order corrections. Moreover, it can be useful either to probe the nucleon structure and improve the knowledge of radiative corrections in QCD. 

In the Regge theory, diffractive processes are described in terms of the exchange of a (soft) Pomeron with vacuum quantum numbers \cite{Collins}. Nevertheless, the nature of the Pomeron and its reaction mechanisms is still not completely known. Since the first measurements on diffractive jets in hadronic collisions, it is usual consider the hard diffractive processes using the diffractive factorization formalism where the hard scattering  resolves the quark and gluon content in the Pomeron \cite{IS} (the so called QCD Pomeron). Systematic observations of diffractive deep inelastic scattering (DDIS) at HERA have increased the knowledge about the QCD Pomeron, providing us with the diffractive distributions of singlet quarks and gluons in the Pomeron as well as the diffractive structure function \cite{H1diff}. In hadronic collisions, a single diffractive event is characterized if one of the colliding hadrons emits a Pomeron that scatters off the other hadron. Hard diffractive events with a large momentum transfer are also characterized by the absence of hadronic energy in certain angular regions of the final state phase space (rapidity gaps). 

Our goal in this work is to estimate the single diffractive cross section for heavy quarks production in heavy ion collisions at the LHC.  Specifically, we have that the coherent diffractive  production of heavy quarks in $AA$ collisions is the process $A+A\rightarrow A+[LRG]+Q\bar{Q}+X$, where $[LRG]$ stands for Large Rapidity Gap. This kind of process exhibits a stronger dependence on energy and atomic number, with the diffractive amplitude proportional to the square of the inelastic one. Therefore, it can serve as a sensitive probe of the low-$x$ dynamics of the nuclear matter.  The incoherent diffractive scattering in $AA$ collisions is the process $A+A\rightarrow A^{*}+[LRG]+Q\bar{Q}+X$, where $A^{*}$ denotes the excited nucleus that subsequently decays into a system of colorless protons, neutrons and nuclei debris. It measures fluctuations of the nuclear color field. Such processes can be investigated experimentally since heavy flavours are copiously produced at LHC energies. For instance, ALICE experiment \cite{ALICE} is very well suited to perform these analyses because it exploits both electron, muon and hadronic channels, having a large rapidity coverage and access to low-$p_T$ region. Moreover, it presents excellent tracking and vertexing capabilities as well as complementary particle identification (TPC,TRD and TOF). Charm production can be measured using exclusive hadronic channels (as $D^0\rightarrow K\pi$) or using semi-exclusive leptonic channels (as $c\rightarrow \ell+X$), whereas bottom can use similar channels, e.g. $B\rightarrow e/\mu+X$  and $b\rightarrow \ell+X$. In special, the tracking capabilities at very low transverse momenta in conjunction with very nice particle identification of ALICE stand it out as one of the most promissing experiments in the physics program of diffractive and electromagnetic reactions \cite{Schicker}.

In the present calculations, we start by the hard diffractive factorization, where the diffractive cross section is the convolution of the diffractive parton distribution functions and the corresponding diffractive coefficient functions, in a similar way as for the inclusive case. However, at high energies there are important contributions from unitarization effects to the single-Pomeron exchange cross section. These absorptive (unitarity) corrections cause the suppression of any large rapidity gap process being  important for the reliability of predictions \cite{k3p}.  The multi-Pomeron contributions depend in general on the particular hard process. They can be estimated by a factor (which is energy and model dependent) called survival probability factor \cite{chehime}.  At the Tevatron energy, $\sqrt s = 1.8$~TeV, the suppression for single diffractive processes is of order 0.05--0.2~\cite{GLM,KMRsoft,KKMR,Prygarin}, whereas at the LHC energy for $pp$ mode, $\sqrt{s}=14$ TeV, the suppression  appears to be 0.06--0.1 ~\cite{GLM,KMRsoft,KKMR,Prygarin}.  There is a large theoretical uncertainty for the nuclear version of suppression factors. We will try to do an educated guess of the suppression factor for coherent and incoherent channels.

This paper is organized as follows. In the next section we summarize the main formulas considered to compute the diffractive ratios for the hadroproduction of charm and bottom. We also present the procedure to estimate the nuclear incoherent and coherent cross sections at the LHC energies. In the last section, we show the numerical results for the inclusive and diffractive cross sections as a function of energy and give predictions for the corresponding diffractive ratios for $pp$ and $AA$ collisions. Discussion on the nuclear dependence of cross sections and corresponding suppression factors are  addressed as well as a comparison to photonuclear reactions in $AA$ collisions is performed.

%
%
\section{Heavy-quark production}

Let us present the main formulas for the inclusive and single diffractive differential cross sections for the production of heavy quarks in proton-proton collisions at high energies. For the diffractive process, the calculation is based on the Ingelman-Schlein (IS) model for diffractive hard scattering \cite{IS}. Accordingly, we will take into account absorption effects by multiplying the diffractive cross section by a gap survival probability factor. Two processes are responsible to heavy quarks production at the Leading Order (LO) in perturbation theory: quark annihilation ($q\bar{q}\rightarrow Q\bar{Q}$), where the pair is always in a colour-octet sate, and gluon fusion ($gg\rightarrow Q\bar{Q}$), where both colour-singlet and octet are allowed. At Next-To-Leading Order (NLO), the $qg+\bar{q}g$ scattering is also included. 

In the inclusive case, the process is described for partons of two hadrons, $h_{a}$ and $h_{b}$, interacting to produce a heavy quark pair, $h_{a}+h_{b}\rightarrow Q\bar{Q}+X$, with center of mass energy $\sqrt{s}$. At LHC energies, the gluon fusion channel dominates over the $q\bar{q}$ annihilation process and $qg$ scattering. The corresponding signal is the production of a $c\bar{c}$ or $b\bar{b}$. The NLO cross section is obtained by convoluting the partonic cross section with the parton distribution function (PDF), $g(x,\mu_F)$, in the proton, where $\mu_F$ is the factorization scale. At any order, the partonic cross section may be expressed in terms of dimensionless scaling functions $f^{k,l}_{ij}$ that depend only on the variable $\rho$ \cite{13magno},
\begin{eqnarray}
\hat{\sigma}_{ij}(\hat{s},m^{2}_{Q},\mu^{2}_{F},\mu^{2}_{R}) & = & \frac{\alpha^{2}_{s}(\mu_{R})}{m^{2}_{Q}}\sum_{k=0}^{\infty}\left [ 4\pi\alpha_{s}(\mu_{R})\right ] ^{k}\\ \nonumber & \times & \sum^{a}_{l=0}f^{(k,l)}_{ij}(\rho)\ln^{l}\left ( \frac{\mu^{2}_{F}}{m^{2}_{Q}}\right )\,
\label{equacao1}
\end{eqnarray}
where $\rho=\frac{\hat{s}}{4m^{2}_{Q}-1}$, $i,j=q,\bar{q},g$, specifying the types of the annihilating partons, $\hat{s}$ is the partonic center of mass, $m_{Q}$ is the heavy quark mass, $\mu_{R}$ is the renormalization scale. It is calculated as an expansion in powers of $\alpha_{s}$ with $k=0$ corresponding to the Born cross section at order ${\cal O}(\alpha^{2}_{s})$. The first correction, $k=1$, corresponds to the NLO cross section at ${\cal O}(\alpha^{3}_{s})$. The dimensionless functions $f_{ij}$ have the following perturbative expansion \cite{nason}
\begin{eqnarray}
f_{ij}\left(\rho,\frac{\mu^{2}}{m^{2}}\right) & = & f^{(0)}_{ij}(\rho) +  g^{2}\left [ f^{(1)}_{ij}(\rho)+\bar{f}^{(1)}_{ij}(\rho)\ln\left(\frac{\mu^{2}}{m^{2}}\right)\right ] \nonumber \\
& + & O(g^{4})\, .
\label{equacao9nason}
\end{eqnarray}
To calculate the $f_{ij}$ in perturbation theory, both renormalisation and factorisation scale of mass singularities must be performed. The subtractions required for renormalisation and factorisation are done at mass scale $\mu$. For instance, since the gluon fusion domains at high energies, the $f^{0}_{gg}(\rho)$ function is given by \cite{nason}
\begin{eqnarray}
f^{(0)}_{gg}(\rho)=\varepsilon\left [\frac{1}{\beta}(\rho^{2}+16\rho+16)\ln\left (\frac{1+\beta}{1-\beta}\right )-28-31\rho\right ]\, ,
\label{equacao15nason}
\end{eqnarray}
where $\varepsilon=\frac{\pi\beta\rho}{192}$. The $f^{(1)}_{ij}(\rho)$ and $\bar{f}^{(1)}_{ij}(\rho)$ functions can be found in \cite{nason}.

The running of the coupling constant $\alpha_{s}$ is determined by the renormalisation group,
\begin{eqnarray}
\frac{d\alpha_{S}(\mu^{2})}{d\ln (\mu^{2})}=-b_{0}\alpha^{2}_{S}-b1\alpha^{3}_{S}+O(\alpha^{4}_{S}),\\   
b_{0}=\frac{33-2n_{1f}}{12\pi}, \,\,\,\,\,\,\,\,\, b_{1}=\frac{153-19n_{1f}}{24\pi^{2}}
\label{equacao11nason}
\end{eqnarray}
where $\alpha_{S}=\frac{g^{2}}{4\pi}$ and $n_{1f}$ is the number of light flavors, $3$ ($4$) to charm (bottom).

The total hadronic cross section for the heavy quark production is obtained by convoluting the total partonic cross section with the parton distribution functions of the initial hadrons \cite{nason}
\begin{eqnarray}\nonumber
\sigma_{ab}(s,m^{2}_{Q}) & = & \sum_{i,j}\int^{1}_{\tau}dx_{1}\int^{1}_{\frac{\tau}{x_{1}}}dx_{2}f^{a}_{i}(x_{1},\mu^{2}_{F})f^{b}_{j}(x_{2},\mu^{2}_{F}) \\  & \times  & \hat{\sigma}_{ij}(\hat{s},m^{2}_{Q},\mu^{2}_{F},\mu^{2}_{R}),
\label{Matrix}
\end{eqnarray}
with the sum $i,j$ over all massless partons. Here, $x_{1,2}$ are the hadron momentum fractions carried by the interacting partons, $f^{a(b)}_{i(j)}$ are the parton distribution functions, evaluated at the factorization scale and assumed to be equal to the renormalization scale in our calculations. We checked the size of NLO corrections by numerically calculating the total inclusive cross sections for heavy flavors pair production using the MRST 2001 LO and MRST 2001 NLO set of partons \cite{14magno}. It was verified that NLO calculations in $pp$ collisions are about a factor $1.5$ greater than the LO calculation, showing the importance of the corrections in NLO. The cross sections were calculated with the following mass and scale parameters: $\mu_{c}=2m_{c}$, $m_{c}=1.5$ GeV, $\mu_{b}=m_{b}=4.5$ GeV, based on the current phenomenology for heavy quark hadroproduction \cite{15magno}.

For the hard diffractive processes, we will consider the IS picture \cite{IS}, where the Pomeron structure (quark and gluon content) is probed. To the single diffraction case, it consists of three steps: first a hard Pomeron is emitted from one of the protons in a small squared four-momentum transfer $|t|$. That hadron is detected in the final state, and the remaining hadron scatters off the emitted Pomeron. Partons from the Pomeron interact with partons from the other hadron and finally, heavy quarks are produced in the final state, from the point-like $Q\bar{Q}$ by the soft gluon radiation. The reaction for heavy quarks hadroproduction is $p+p\rightarrow p+Q\bar{Q}+X$. In this approach, the single diffractive cross section is assumed to factorise into the total Pomeron-hadron cross section and the Pomeron flux factor \cite{IS}. The single diffractive event may then be written as \cite{magno}
\begin{eqnarray}
 \label{sdexp}
\frac{d\sigma^{\mathrm{SD}}\,(hh\rightarrow h+  Q\bar{Q}+X)}
{dx^{(a)}_{\pom}d|t_a|}\! & = & \! f_{{\rm\pom}/a}(x^{(a)}_{\pom},|t_a|)\\ \nonumber  & \times & \sigma\left({\pom} + h\rightarrow  Q\bar{Q}  +  X\right),\nonumber
\end{eqnarray}
where $x_{\pom}$ is the Pomeron kinematical variable, defined as $x^{a}_{\pom}=s^{(b)}_{\pom}/s_{ab}$, where $\sqrt{s^{b}_{\pom}}$ is the center-of-mass energy in the Pomeron-hadron $b$ system and $\sqrt{s}_{ab}=\sqrt{s}$ is the center-of-mass energy in the hadron$_{a}$-hadron$_{b}$ system, with $t_{a}$ denoting the momentum transfer in the hadron $a$ vertex.

To obtain the corresponding expression for hard diffractive processes, one assumes that one of the hadrons, lets say a hadron $a$, emits a Pomeron whose partons interact with partons of the hadron $b$. So, the parton distribution in Eq. (\ref{Matrix}) is replaced by the convolution between a  distribution of partons in the Pomeron, $\beta f_{a/\pom}(\beta,\mu^{2})$, and the "emission rate" of Pomerons by the hadron, $f_{\pom/h}(x_{\pom},t)$, called the Pomeron flux factor. Its explicit formulation is described in terms of Regge theory, and so, the expression for the single diffractive cross section for $Q\bar{Q}$ production is written as \cite{magno}
\begin{eqnarray}
\label{sdxsect}
& & \sigma_{ab}^{\mathrm{SD}}(s,m_Q^2)  =   \sum_{i,j=q\bar{q},g} 
\int_{\rho}^1dx_1\int_{\rho/x_1}^1 dx_2 \int_{x_1}^{x_{\pom}^{\mathrm{max}}}\frac{dx_{\pom}^{(1)}}{x_{\pom}^{(1)}} \nonumber \\ \nonumber & \times & \bar{f}_{\pom/a}\left(x_{\pom}^{(1)}\right)f_{i/\pom}\left(\frac{x_1}{x_{\pom}^{(1)}},\mu^2\right)  f_{j/b}(x_2,\mu^2)\,
\hat \sigma_{ij}(\hat{s} ,m_Q^2,\mu^2) \\  & + & \, (1\rightleftharpoons 2)\,.
\end{eqnarray}

For the heavy quarks production in nucleus-nucleus collisions, two processes can occur: the first one is a coherent process, which is when a nucleus emits a Pomeron and partons of that Pomeron interact with partons from the another nucleus. In this case, the calculations are carried out in similar way as described above, where the proton form-factor,
\begin{eqnarray}
F_{1}(t)=\frac{4m^{2}_{p}-2.8t}{4m^{2}_{p}-t}\left ( 1-\frac{t}{0.7\,GeV^{2}}\right ) ^{-2}\, ,
\label{equacao3agababyan}
\end{eqnarray}
is replaced by the nucleus form-factor which is parameterized as \cite{agababyan}
\begin{eqnarray}
F_A(t)\approx \exp\left(\frac{R^{2}_{A}t}{6}\right)\, ,
\label{equacao16agababyan}
\end{eqnarray}
where $R_{A}$ is the radius of the nucleus $A$ ($R_{A}=1.2\,A^{1/3}$ fm). That is, the nucleus-Pomeron coupling has the form $\beta_{A\pom}\propto A\beta_0|F_A(t)|$, where $\beta_0$ is the quark-Pomeron coupling. Here, we assume the Donnachie-Landshoff notation for the nucleon-Pomeron coupling, $\beta_{N\pom}=3\beta_0F_1(t)$, and then for the nucleus-Pomeron interaction it is used the additivity of the total nucleon-nucleon cross section and replace $\beta_0$ by $A\beta_0$ and the isoscalar magnetic nucleon form factor $F_1(t)$ by the elastic nuclear form factor $F_A(t)$. This procedure has been used for a long time to compute the Pomeron-Pomeron contribution in heavy ion collisions \cite{Muller,Natale}.  Notice that a linear dependence of the nucleus-Pomeron coupling on $A$ is strong. However, it is a reliable approximation for the hard process considered here.
On other hand, the incoherent process is characterized by an emission of the Pomeron for one of the protons in the nucleus. Studies for the $A$-dependence of the incoherent diffractive scattering of symmetric nuclei allows the cross section to be parameterized as \cite{agababyan}. In the impulse approximation, this process can be described as the interaction on a nucleus with a Pomeron belonging to a nucleon embedded in the nuclear medium of the remaining nucleus. Thus, we have a roughly $A^2$ dependence for the incoherent cross section:
\begin{eqnarray}
\sigma_{A}^{\mathrm{inc}}\approx A^{2\,\alpha}\sigma_{N}
\label{equacao17agababyan}
\end{eqnarray}
where $\alpha=1$ in our case. In the literature, some authors use $\alpha =0.7-0.8$ for diffractive process \cite{38agababyan} and $\sigma_{N}$ is the diffraction dissociation nucleon-nucleon cross section. This theoretical parameterization for the nuclear dependence is based on the Regge approach to strong interactions at high energies. The energies for nucleus-nucleus interactions considered here are $\sqrt{s}=5.5$ TeV for PbPb beams and $\sqrt{s}=6.3$ TeV for CaCa collisions. 

The procedure to compute the hard nuclear cross section above is still arbitrary. Despite soft diffraction in nuclear targets to be reasonably described using Glauber theory, the situation is not clear concerning hard diffraction in nuclear collisions. As we are considering the Ingelman-Schlein Pomeron, there are two possibilities: (a) one defines a new Pomeron flux depending on the atomic number and differential cross section normalized by the Pomeron-nucleon cross section, $\sigma_{\pom p}$,  or (b) one defines a sort of Pomeron-nucleus cross section, $\sigma_{\pom A}$, keeping the original Pomeron flux unchanged. These issues were first addressed in Ref. \cite{Covolan}, where an analysis of diffractive dissociation of nuclei in proton-nucleus and meson-nucleus was presented. Here, we have chosen the option (a) due to its simplicity of implementation. We also assume no gluon shadowing in the interacting nucleus. The introduction of shadowing in the diffractive cross section should reduce it by 60-70 $\%$ as $R_g(x\simeq 10^{-4})\approx 0.65$ \cite{EKS}, where roughly speaking $x\simeq m_Q/\sqrt{s}$. The effect in the diffractive ratio is weaker and probably increase it as the new inclusive cross section is smaller than the minimum bias approximation.

In the estimates for the cross sections in Eq. (\ref{sdxsect}), we consider a standard Pomeron flux from Regge phenomenology and which is constrained from the experimental analysis of the diffractive structure function \cite{H1diff}. For the diffractive gluon distribution in the Pomeron, $g_{\pom}(x_1,\mu_F^2)$, we will consider the diffractive PDFs obtained by the H1 Collaboration at DESY-HERA \cite{H1diff}, where the Pomeron structure function has been modeled in terms of a light flavor singlet distribution $\Sigma(x)$, i. e., the $u$, $d$ and $s$ quarks with their respective anti-quarks. Also, it has a gluon distribution $g(z)$, with $z$ being the longitudinal momentum fraction of the parton entering the hard sub-process with respect to the diffractive exchange.  In our numerical calculations, it will be the cuts for the integration over $x_{\pom}$,  $x_{\pom}^{min} = 0.05$.

As a final step in our estimates of the single diffractive cross section, we will consider the suppression of the hard diffractive cross section by multiple-Pomeron scattering effects (absorptive corrections). This is taken into account through a gap survival probability $<|S|^{2}>$, which can be described in terms of screening or absorptive corrections \cite{Bj}.  There are intense theoretical investigations on this subject in the last years. We quote Ref. \cite{Prygarin} for a discussion and comparison of theoretical estimations for the gap survival probabilities. We notice that it is the main theoretical uncertainty in the present calculation of diffractive ratios. As a baseline value, we follow  Ref. \cite{KKMR}, which considers a two-channel eikonal model that embodies pion-loop insertions in the Pomeron trajectory and high mass diffractive dissociation. For LHC energy on $pp$ collisions, one has $<|S|^{2}>=0.06$. In single channel eikonal models, this factor can reach up to $0.081-0.086$ as discussed in Ref. \cite{Prygarin}.  

Concerning the model dependence, the single channel eikonal model considers only elastic re-scatterings, whereas for the multi channel one takes into account also inelastic diffractive intermediate re-scatterings.  The available experimental observables which can be compared to the theoretical predictions of the survival probability factor are the hard LRG di-jets data obtained in the Tevatron and HERA \cite{Prygarin,KKMR}  as well as diffractive hadroproduction of heavy bosons ($W^{\pm}$ and $Z^0$) in the Tevatron \cite{GDMM}. Here, some discussion for the nuclear case is in order. Currently, there is a lack of information on the gap survival probability in nucleus-nucleus and proton-nucleus collisions. It is known that it will be much smaller than for proton-proton collisions because of multiple collisions of projectile nucleons.  In Ref. \cite{JMiller}, the central exclusive diffraction Higgs production in collisions with nuclei is considered, where gluon-gluon fusion even in proton-nucleus collisions leads to a very small cross section, while $\gamma \gamma$ fusion gives the dominant contribution. The conclusion is that the value of the survival probabilities is negligible
small for gluon-gluon fusion. The calculation gives $<|S|^{2}>_{gg\rightarrow H } = 8 \times 10^{-4}$ for proton-ion collisions at the LHC energy and even  small ( $<|S|^{2}>_{gg\rightarrow H } \approx 8.1 \times 10^{-7}$ ) for ion-ion collisions.

\subsection{Results and discussion}

Let us present the results for the inclusive and diffractive heavy quarks cross  sections for hadronic and nucleus-nucleus collisions. The calculations for the inclusive and diffractive cross sections, as well the diffractive ratios to heavy quark production in proton-proton collisions are showed at Tab. (\ref{tabelahadronica}). We take the value $<|S|^{2}>_{pp}=0.06$ for the absorption corrections in hadronic collisions at the LHC. The partons PDF and scales are mentioned in previous section. For the diffractive gluon PDF, we take the experimental FIT A (the fully integrated cross section is insensitive to a different choice, i.e. FIT B). The main theoretical uncertainty in the diffractive ratio is the survival probability factor, whereas uncertainties associated to factorization/renormalization scale, parton PDFs and quark mass are minimized taking a ratio. The present results are consistent with a recent estimation computed in Ref. \cite{magno}, where a value $<|S|^{2}>=0.09$ was considered. In that work, a LO version of cross sections was considered and then it can be verified that even the NLO corrections are absorbed in a ratio.

\begin{table}[t]
\centering
\renewcommand{\arraystretch}{1.5}
\begin{tabular}{|c|c|c|c|}
\hline
	Heavy Quark       &   $\sigma_{\mathrm{inc}}(\sqrt{s}=14\, \mathrm{TeV})$    &	$\sigma_{\mathrm{diff}}(\sqrt{s}=14\, \mathrm{TeV})$	&	$R_{\mathrm{diff}}$ \\\hline
$c\bar{c}$	&	7811    $[\mu b]$        &  		178  $[\mu b]$     	&		2.3 \%      \\ \hline 
$b\bar{b}$	&	393    $[\mu b]$  	    &	    	7   $[\mu b]$      	&		1.7	\%    \\\hline
\end{tabular}  	
\caption{The inclusive and single diffractive (corrected by absorption effects) cross sections in $pp$ collisions at the LHC. The corresponding diffractive ratios, $R_{\mathrm{diff}}$, are also presented.}
\label{tabelahadronica}
\end{table}

\begin{table*}[t]
\centering
\renewcommand{\arraystretch}{1.5}
\begin{tabular}{|c|c|c|c|c|c|}
\hline
            Incoherent              &     CaCa  ($c\bar{c}$) &    PbPb ($c\bar{c}$)  &    CaCa ($b\bar{b}$)   &   PbPb ($b\bar{b}$)  \\ \hline
$\sigma_{\mathrm{inc}}/A^2$                  &    $1.94$ mb             &   $1.68$ mb           & $0.04$ mb               &   $0.03$ mb           \\ \hline
$\sigma_{\mathrm{inc}}^{\mathrm{abs}}$              &    $186-0.003$ mb             &   $4356-0.07 $ mb            & $3.78-6.3\times 10^{-5}$  mb              &   $85-0.001$ mb           \\ \hline
$R_{\mathrm{inc}} [\%]$                         &    $40$              &   $38$                & $20$                &   $19$            \\ \hline
$R_{\mathrm{inc}}^{\mathrm{abs}} [\%]$          &    $2.4-4\times 10^{-5}$              &   $2.28-3.8\times 10^{-5}$          & $1.2-2\times 10^{-5}$                 &   $1.14-1.9\times 10^{-5}$  	      \\ \hline
\end{tabular}  	
\caption{The incoherent cross section per nucleon for calcium and lead without absorptive corrections. The cross section including absorptive corrections ($\sigma_{\mathrm{inc}}^{\mathrm{abs}}$) and the diffractive ratios are also presented (see text).}
\label{tabelaIncoerente}
\end{table*}

As referred before, in nuclear collisions there are not calculations of $<|S|>^{2}$ for single diffraction processes. Estimations of central Higgs production in $pA$ and $AA$ collisions give values much lower than $10^{-4}$ \cite{JMiller}. Thus, for sake of illustration we calculated the diffractive cross sections and the diffractive ratios in nuclear collisions using a theoretical error band for GSP: the upper value is obtained using survival probability for $pp$ collisions at 14 TeV  and the lower value is obtained using the estimation from Ref. \cite{JMiller}, $<|S|>^{2}_{AA}=10^{-6}$. The results are shown in Tabs. (\ref{tabelaIncoerente}) and (\ref{tabelaCoerente}) for incoherent  and coherent  collisions, respectively. The ratios are obtained from 
\begin{eqnarray}
R_{\mathrm{inc}}=\frac{\sigma_{A}^{inc}}{\sigma_{A}},\,\,\,\,\,\,R_{\mathrm{coh}}=\frac{\sigma_{A}^{coh}}{\sigma_{A}}
\label{equacaoIC}
\end{eqnarray}
where $\sigma_{A}^{inc}$ is given by Eq. (\ref{equacao17agababyan}) and $\sigma_{A}=A^{2}\sigma_{N}$. For the coherent case,  $\sigma_A^{coh}$ being the $\sigma_{N}$ cross section with the modification in text below Eq. (\ref{equacao16agababyan}). As a cross check, we found the following values for the inclusive total cross sections per nucleon: 4.8 (0.2) mb for charm (bottom) production in CaCa collisions and  4.3 (0.17) mb for PbPb collisions at the LHC.

\begin{table*}[t]
\centering
\renewcommand{\arraystretch}{1.5}
\begin{tabular}{|c|c|c|c|c|c|}
\hline
  Coherent                          &   CaCa  ($c\bar{c}$)  &      PbPb ($c\bar{c}$) &    CaCa ($b\bar{b}$)&   PbPb ($b\bar{b}$)  \\ \hline
$\sigma_{\mathrm{coh}}/A^2 $                 &	$2.9$ mb                &  $3.7$ mb                & $0.05$ mb           &  $0.06$ mb  \\ \hline
$\sigma_{\mathrm{coh}}^{\mathrm{abs}}$               &	$277-5\times 10^{-3}$  mb              &  $9686-0.16$  mb               & $5.2-8.6\times 10^{-5}$ mb             &  $156-0.003$ mb \\ \hline
$R_{\mathrm{coh}} [\%]$                         &    $60$              &   $86$                & $27$                &   $35$            \\ \hline
$R_{\mathrm{coh}}^{\mathrm{abs}} [\%]$          &    $3.6-6\times 10^{-5}$              &   $5.2-8.6\times 10^{-5}$          & $1.6-2.7\times 10^{-5}$                 &   $2.1-3.5\times 10^{-5}$  	      \\ \hline
\end{tabular}  	
\caption{The coherent cross section  per nucleon for calcium and lead without absorptive corrections. The cross section including absorptive corrections ($\sigma_{\mathrm{coh}}^{\mathrm{abs}}$) and the diffractive ratios are also presented (see text).}
\label{tabelaCoerente}
\end{table*}

In Tab. (\ref{tabelaIncoerente}), the incoherent cross section per nucleon, $\sigma_{\mathrm{inc}}/A^2$, is presented for calcium and lead without absorptive corrections. The cross section including absorptive corrections, $\sigma_{\mathrm{inc}}^{\mathrm{abs}}$, is shown as a band: the upper value is obtained using the gap survival probability $<|S|^{2}>_{pp}(\sqrt{s}=14 \,\mathrm{TeV}) = 0.06$ and the lower one  corresponds to $<|S|^{2}>_{AA} \simeq 10^{-6}$ \cite{JMiller}.  The incoherent cross section including the absorption corrections for the charm case in PbPb collisions has magnitude of dozens of micro-barns in the lower band, which is a hopeful sign from the experimental point of view. For bottom, the situation is similar with a lower bound of order 1 $\mu$b. The diffractive ratios are also presented: $R_{\mathrm{inc}}$  stands for the ratio without considering absorption factor, whereas  $R_{\mathrm{inc}}^{\mathrm{abs}}$ stands for ratios taking absorption (the band is similar as for the diffractive cross section). The ratio without absorptive corrections is almost identical to the diffractive ratios for the proton-proton case. The reason is the $A$-dependence for the incoherent cross section $\sigma_{\mathrm{inc}}^{\mathrm{abs}}\propto A^2\sigma^{\mathrm{SD}}_{pp}$.  The ratio is very small but is still larger than for double diffractive production (double Pomeron exchange), as computed in Ref. \cite{agababyan}.

The coherent cross section and ratios are presented in  Tab. (\ref{tabelaCoerente}), using the same notation as for the incoherent case. It is noticed that the coherent cross section is larger than the incoherent one, reaching a factor 2 for PbPb collisions. This enhancement is translated to the diffractive ratios as well.  The reason is the $A$-dependence for the coherent cross section $\sigma_{\mathrm{coh}}\propto A^{\alpha}\sigma^{\mathrm{SD}}_{pp}$,  where $\alpha = 7/3$.  The coherent cross section including the absorption corrections for the charm case in PbPb collisions reaches  200  $\mu$b in the lower band, which is still  hopeful, and for bottom one gets a lower bound of order 3 $\mu$b.

Let us now comment on the $A$-dependence of coherent and incoherent scattering. For the incoherent case, we consider the impulse approximation where a Pomeron is emitted from one nucleon embedded in one of the incoming nucleus, This Pomeron then interacts with the remaining nucleus, where we have assumed no gluon shadowing $xg_A(x_2,\mu_Q)=A\,xg_p(x_2,\mu_Q)$. Such approximation gives $\sigma_{\mathrm{inc}}^{AA}\propto A^{2}\,\sigma^{\mathrm{SD}}_{pp}$. If we consider $pA$ collisions, the the Pomeron would interact with a proton and we would get  $\sigma_{\mathrm{inc}}^{pA}\propto A\,\sigma^{\mathrm{SD}}_{pp}$. In the coherent case, the $A$ dependence comes from the Pomeron flux for as nucleus, $f_{\pom/A}(x_{\pom},t)\propto A^2|F_A(t)|^2$ (the Pomeron is emitted by the nucleus as a whole). After integration over $t$, one gets  $\bar{f}_{\pom/A}(x_{\pom})\propto A^2/R_A^2$, with $R_A^2=1.44\, A^{2/3}$ fm$^2$. Once again supposing no gluon shadowing in the remaining nucleus, one has $\sigma_{\mathrm{coh}}^{AA}\propto A^{7/3}\,\sigma^{\mathrm{SD}}_{pp}$. Using the same arguments it is easy to see that in $pA$ collisions we would find $\sigma_{\mathrm{coh}}^{pA}\propto A^{4/3}\,\sigma^{\mathrm{SD}}_{pp}$ (the $A$-dependence comes only from the integrated Pomeron flux). Notice that for $pA$ collisions our calculation is in agreement with the Born approximation for the interaction with a single nucleon in the Glauber-Gribov approach. Moreover, we verify a parametric enhancement by a factor $A^{1/3}$ of the coherent diffractive cross section with respect to the incoherent one in both $pA$ and $AA$ collisions. This fact explains the distinct values verified in Table (III) compared to Table (II) and the reason for a different enhancement for CaCa and PbPb. Such a similar enhancement has been recently found in Ref. \cite{Tuchin}, where the cross section for incoherent and coherent diffractive gluon production in $q\bar{q}A$ collisions (considered a prototype of $pA$ scattering at the LHC) was computed in the Color Glass Condensate framework. 

Concerning the main theoretical uncertainty on the present calculation for the diffractive ratios, some words of caution are in order. Of course, the error bands in Tables (II) and (III) are a naive oversimplification. The suppression factors in nuclear collusions are well known to depend on atomic number, center of mass energy and on the specific produced final state \cite{Martin}. For instance, in Ref. \cite{Martin} a careful calculation of absorption corrections for proton-nucleus collisions was carried out considering Drell-Yan and heavy quark production (charm and bottom). It was shown that the suppression takes distinct values for coherent and incoherent diffraction, being the absorption on the coherent case one order of magnitude stronger than the incoherent one. For instance, using the single-channel Glauber approach (which is insensitive to the parton longitudinal momentum fraction) the prediction gives a factor $4\times 10^{-2}$ for charm/bottom production in  incoherent scattering for proton-gold collisions at $\sqrt{s}=300$ GeV (RHIC kinematics) and a factor  $5\times 10^{-3}$ for the corresponding coherent scattering. As the dependence of the absorption factors is mild on energy, the estimation presented in Ref. \cite{JMiller} for the suppression in exclusive central diffraction in $pA$ collisions ($<|S|^{2}>_{gg\rightarrow H } = 8 \times 10^{-4}$) is either reasonable. The situation is however unclear for the nucleus-nucleus collisions. Hopefully, we can try to do an educated guess for the absorption factor in single diffraction cross section in heavy ion collisions. In order to do so, we will use the procedure presented in Ref. \cite{Pajares}, where the central diffraction and single diffraction cross sections in nucleus-nucleus collisions are computed using the so-called {\it criterion C} (we quote Ref. \cite{Pajares} for further details). The single diffraction (coherent) cross section for $AB$ collisions is given by:
\begin{eqnarray}
\sigma^{\mathrm{SD}}_{AB}  =  \sigma^{in}_{AB}\left(\sigma^{in}_{pp}\right)-\sigma^ {in}_{AB}\left(\sigma^{in}_{pp}-\sigma^{\mathrm{SD}}_{pp}\right),
\label{sigsdaa}
\end{eqnarray}
where $\sigma^{in}_{AB}$ is the inelastic $AB$ cross section considered as a function of the nucleon-nucleon total cross section $\sigma$ ($\sigma^{in}_{pp}$ and $\sigma^{\mathrm{SD}}_{pp}$ are the inelastic and single diffractive cross sections in proton-proton case, respectively). The expression in the equation above can be explicitly written for $pA$ collisions where the dependence on the inelastic cross section, $\sigma^{in}_{pA}(\sigma )=1-[1-\sigma\, T(b)]^A \simeq 1-\exp [-A\,\sigma T(b)]$, is known in Glauber model for fixed impact parameter $b$. Therefore, using such an information and taking $\sigma^{\mathrm{SD}}_{pp}$ very small in Eq. (\ref{sigsdaa}), one obtains $\sigma^{\mathrm{SD}}_{pA}=A_{\mathrm{eff}}\times \sigma^{\mathrm{SD}}_{pp}$, with $A_{\mathrm{eff}}=A\int d^2b \,T_A(b)\,\exp[-A\,\sigma^{in}_{pp}\,T_A(b)]$. The application of the formalism to $AA$ collisions turns out to be difficult due to the absence of an explicit expression for $\sigma^{in}_{AA}(\sigma)$. In Ref. \cite{Pajares} the authors considered optical approximation in which $\sigma^{in}_{AA}$ is given by corresponding formula for $pA$ with $A\rightarrow AB$ and an effective profile function for two colliding nuclei, $T_{AB}=\int d^2\bar{b}\,T_A(\bar{b})\,T_B(b-\bar{b})$. The final expression for single diffractive (coherent)  cross section in $AA$ collisions is given by
\begin{eqnarray}
& & \sigma^{\mathrm{SD}}_{AA} \,(\sqrt{s};\,A)  =  A_{\mathrm{eff}}^2 \times \sigma^{\mathrm{SD}}_{pp}(\sqrt{s}),\,\label{sdxs}\\
& & A_{\mathrm{eff}}^2  =  A^2\int d^2b \,T_{AA}(b)\,\exp\left[-A^2\,\sigma^{in}_{pp}\,T_{AA}(b)\right].
\label{sdxs1}
\end{eqnarray} 

Using Woods-Saxon nuclear densities and considering the inelastic cross section $\sigma^{in}_{pp}\,(\sqrt{s}=6\,\mathrm{TeV})=73$ mb \cite{Pajares} the values for the effective atomic number of the colliding nuclei are $ A_{\mathrm{eff}}^2  =  6.21$ for calcium and $ A_{\mathrm{eff}}^2  = 9.52$ for lead nucleus.

Numerically, the cross sections obtained for the LHC using Eqs. (\ref{sdxs}-\ref{sdxs1}) are the following: $\sigma^{\mathrm{SD}}_{\mathrm{PbPb}}= 1.17\,(0.02)$ mb and $\sigma^{\mathrm{SD}}_{\mathrm{CaCa}}= 0.84\,(0.017)$ mb for charm (bottom). In order to obtaining them we have for the proton case $\sigma^{\mathrm{SD}}_{pp}(\sqrt{s}=5.5\,\mathrm{TeV})= 120\,(2)$ $\mu$b and  $\sigma^{\mathrm{SD}}_{pp}(\sqrt{s}=6.3\,\mathrm{TeV})=136 \,(2.8)$ $\mu$b  for charm (bottom), where the single-Pomeron calculation have been corrected by absorption factor in proton-proton collisions. Thus, we considered $<|S|^{2}>^{\mathrm{KMR}}_{pp}=0.073\,(0.07)$ for $\sqrt{s}=5.5 \,(6.3)$ TeV, which is obtained using  a parametric interpolation formula for the KMR survival probability factor in the form $<|S|^{2}>=a/[b+\ln (\sqrt{s/s_0})]$ with $a = 0.308$, $b=-4.42$ and $s_0=1$ GeV$^2$. This formula interpolates between the  (single diffraction) survival probabilities $10 \, \%$ at Tevatron and $6 \, \%$ at the LHC. The single diffractive cross section can be also estimated for $pA$ collisions using similar procedure. It is obtained $\sigma^{\mathrm{SD}}_{\mathrm{pPb}}= 0.76\,(0.018)$ mb for charm (bottom), where $A_{\mathrm{eff}}= 4.39$ \cite{Pajares}. Having the corrected values for the diffraction dissociation cross section, we are now in conditions to estimate the overall suppression factor in coherent case. In order to do so, one takes the ratio between the corrected cross section and the single-Pomeron calculation show in the first row of Table (III). This gives a suppression factor of  $S_A=7.3\times 10^{-6}$ for charm and $S_A=7.7\times 10^{-6}$ for bottom production in coherent single diffraction in PbPb collisions at the LHC. These values are not so far away from the estimation of Ref. \cite{JMiller}, where central diffractive Higgs production is considered. Following Ref. \cite{Martin}, the suppression factor for incoherent diffraction should be one order of magnitude larger than for the coherent case. This fact has an important consequence as it compensates the smaller incoherent cross section. The enhancement of coherent to incoherent is proportional to $A^{1/3}$ as referred before and this gives a factor around ten for lead nucleus.

Finally, we compare the present calculation with the (inclusive) photoproduction of heavy quarks in ultraperipheral collisions (UPC's) in heavy ion collisions \cite{PRep}. The reason is that the final state configuration is similar to the coherent diffraction (both nuclei remain intact and one rapidity gap). For instance, in Refs. \cite{UPC1,UPC2} the inclusive charm and bottom production is computed in coherent heavy ion interaction  considering distinct theoretical approaches for the heavy quark production. It was found $\sigma^{\mathrm{upc}}_{\mathrm{PbPb}}= 633-2079$ mb for charm and $\sigma^{\mathrm{upc}}_{\mathrm{PbPb}}= 8.9-18$ mb for bottom at the LHC, where the lower bound corresponds to the result considering saturation model and the upper limit corresponds to $k_{\perp}$-factorization approach. In $pA$ collisions one has $\sigma^{\mathrm{upc}}_{\mathrm{pPb}}= 5-17$ mb for charm and $\sigma^{\mathrm{upc}}_{\mathrm{pPb}}= 81-155$ $\mu$b for bottom (we quote Ref. \cite{UPC3} for further details). Therefore, for heavy ions the photoproduction channel should dominate over the single diffractive channel. The situation is completely the opposite in the proton-proton case. Recently, the inclusive photoproduction of heavy quarks was computed for $pp$ collisions at the LHC \cite{pp1,pp2} using distinct saturation models, which provides us with the estimations $\sigma^{\mathrm{upc}}_{pp}= 7.54-3.66$ $\mu$b for charm and $\sigma^{\mathrm{upc}}_{pp}= 0.16-0.05$ $\mu$b for bottom. These values can be compared to Table (I) and it is verified that the single diffractive channel dominates over photoproduction channel. The enhancement of UPC's compared to single diffractive channel in $AA$ reactions is easily understood from their distinct $A$-dependences: photoproduction grows as $\propto A^3$ ($Z^2$ enhancement from the equivalent photon flux plus an additional enhancement from the photonuclear cross section, $\sigma_{\gamma A}\propto A\,\sigma_{\gamma p}$), whereas single diffractive cross section gets a factor $A_{\mathrm{eff}}^2\simeq A^{1/3}$ as shown in Eqs. (\ref{sdxs}-\ref{sdxs1}).

In summary, we have presented predictions for diffractive heavy flavor production in heavy ion collisions at the LHC. The cross sections are large enough and can be investigated experimentally.  In our calculations,the Ingelman-Schlein picture for hard diffraction was considered further corrected by absorption corrections given by gap survival probability factor. For the Pomeron structure function, the H1 diffractive parton density functions were considered. We investigate the theoretical uncertainty on the multiple interaction corrections for the nuclear case and addressed the coherent and incoherent scatterings. We are aware of the limitations and theoretical incompleteness of such a picture. However, we think it is reasonable for a first exploratory study. An alternative approach would be consider the nuclear version for the diffractive production of heavy quarks within the light-cone dipole approach of Ref. \cite{Boris}. In that work novel leading twist mechanisms of diffractive excitation of heavy flavors in hadronic collisions are proposed, which broke the factorization leading to higher twist diffraction. Returning to the present investigation, the main results are the estimations for $AA$ and $pA$ collisions: we obtained $\sigma^{\mathrm{SD}}_{\mathrm{PbPb}}= 1.17\,(0.02)$ mb and  $\sigma^{\mathrm{SD}}_{\mathrm{pPb}}= 0.76\,(0.018)$ mb for charm (bottom). As a byproduct, we also estimate the overall suppression factor in coherent diffraction, which reaches $S_A\simeq 7\times 10^{-6}$ for heavy quark production  in PbPb collisions. The corresponding factor for incoherent diffraction would be one order of magnitude larger, which it would compensate the parametric enhancement by a factor $A^{1/3}$ of the coherent diffractive cross section with respect to the incoherent one in both $pA$ and $AA$ collisions. We verified that the single diffractive channel dominates over photoproduction channel in proton-proton case, whereas it is one or two orders of magnitude smaller than photoproduction in heavy-ion collisions. The enhancement of ultraperiphetral collisions  compared to single diffractive channel  is driven by their different $A$-dependences. Notice that the present calculation we provided the fully integrated cross sections and an accurate study using relevant kinematic cuts (rapidity gap separation and transverse momentum spectrum) in the LHC is in order.

%


\section*{Acknowledgments}
The authors are grateful to Rainer Schicker (Heidelberg University and ALICE/CERN) for motivating us to perform the present calculations. Magno Machado thanks useful discussions and remarks from  V. Serbo, A. Szczurek and Wlodek Gurin.  Mairon Machado would like to thank  E. Levin, E. Gotsman and J. Miller for informative discussions. 
This work was supported by CNPq and FAPERGS, Brazil.


%
%


\end{document}